\documentclass[aps,prb,reprint,superscriptaddress]{revtex4-2}
\usepackage{amsmath,amssymb,graphicx}

\begin{document}

\title{Formal $O(N^3)$-Scaling Second-Order Perturbation Theory by Block Tensor Decomposition: Implementation on {MP2} and {rPT2}}

\author{Yueyang Zhang}
\author{Wei Wu}
\author{Peifeng Su}
\email{supi@xmu.edu.cn}
\affiliation{State Key Laboratory of Physical Chemistry of Solid Surfaces,\\
Fujian Provincial Key Laboratory of Theoretical and Computational Chemistry,\\
and College of Chemistry and Chemical Engineering, Xiamen University,\\
Xiamen, Fujian 361005, China}

\date{19 May 2026}

\begin{abstract}
Block tensor decomposition (BTD) and canonical polyadic decomposition (CPD) are combined into a unified $O(N^3)$-scaling framework for second-order perturbation theory (PT2), demonstrated on MP2 and renormalized PT2 (rPT2). BTD constructs the tensor hyper-contraction kernel at $O(N^3)$ via a dual-grid scheme; CPD factorizes the exchange channel through a block-based two-stage ALS. An asymmetric half-kernel design applies bare Coulomb to one vertex and coupling-constant-averaged screening to the other, capturing the SOSEX component of rPT2 without a frequency-dependent CPD. For MP2, BTD-CPD reproduces canonical RI-MP2 to 0.058~kcal/mol per heavy atom. For rPT2@PBE0 on the S66x8 benchmark, the mean absolute error is 0.36~kcal/mol (ME $-$0.19, RMSE 0.46) over 528 data points. The CPD-compressed intermediates yield $O(N^2)$ storage alongside $O(N^3)$ scaling.
\end{abstract}

\maketitle

\section{Introduction}

Wave-function-based electron correlation methods such as canonical coupled-cluster theory~\cite{Cizek1966,Bartlett1981,Bartlett2007,Crawford2000,Shavitt2009,Helgaker2000} deliver chemical accuracy but scale as $O(N^6)$--$O(N^7)$, restricting their routine application to systems of 20--30 atoms. Kohn--Sham density functional theory (DFT) handles hundreds of atoms but its accuracy depends on the quality of the approximate exchange-correlation functional, which lacks a systematic path to higher accuracy. Second-order M{\o}ller--Plesset perturbation theory (MP2)~\cite{Moller1934}, a widely used post-Hartree--Fock method, captures dispersion at $O(N^5)$ cost but diverges for small-gap systems, most notably along bond-dissociation curves, and lacks higher-order screening. Double-hybrid density functionals~\cite{Grimme2006,Schwabe2007,Chai2009,Zhang2009,Goerigk2011,Goerigk2014,Mardirossian2018,Martin2020} mix MP2-like correlation into DFT, improving accuracy but inheriting MP2's gap-dependence. A wavefunction-based method that delivers comparable accuracy without density-functional approximations remains attractive.

A distinct route to taming the MP2 divergence is regularization. Orbital-energy-dependent regularizers such as $\kappa$-MP2~\cite{Lee2018} damp the problematic small-denominator contributions, while size-consistent Brillouin--Wigner PT2 (BW-s2)~\cite{CarterFenk2023,CarterFenk2023b} repartitions the Hamiltonian to shift orbital energy gaps away from zero. Both approaches restore finite correlation energies at bond dissociation while preserving the $O(N^5)$ scaling of conventional MP2.

The random phase approximation (RPA)~\cite{Bohm1953,Langreth1977,Furche2001} achieves regularization through a different mechanism: by summing all ring diagrams to infinite order, it naturally avoids the divergence without empirical parameters or gap-shifting operators. RPA occupies the highest rung of Jacob's ladder~\cite{Perdew2001}, depending on unoccupied orbitals and their eigenvalues beyond the exact-exchange component available to hybrid functionals. Derived from the adiabatic-connection fluctuation-dissipation theorem (ACFDT)~\cite{Langreth1977,Gunnarsson1976}, RPA captures long-range dispersion and, when built on exact exchange, eliminates one-electron self-interaction error~\cite{Eshuis2012,Ren2012a}. Its main drawback is the same-spin self-correlation error: RPA assigns a finite correlation energy to same-spin electron pairs, where Pauli exclusion requires zero. This produces systematic underbinding~\cite{Gruneis2009,Paier2010}. Spin-component scaling~\cite{Zhang2019} and range separation~\cite{Toulouse2009} offer alternative corrections. The second-order screened exchange (SOSEX) corrects this by evaluating a single exchange diagram with RPA-screened amplitudes~\cite{Freeman1977,Gruneis2009,Jansen2010}. For non-Hartree--Fock references (e.g., GGA or hybrid functionals), Brillouin's theorem breaks down and single excitations contribute at first order; the renormalized single-excitation (rSE) term recovers this contribution~\cite{Ren2011}. Together, RPA, SOSEX, and rSE form the renormalized second-order perturbation theory (rPT2)~\cite{Ren2013}, which delivers well-balanced accuracy for thermochemistry, reaction barriers, and non-covalent interactions without empirical parameters. However, rPT2 inherits the $O(N^5)$ scaling of conventional post-Hartree--Fock methods. An $O(N^3)$ variant would reach large organic molecules, molecular crystals, and biomolecular fragments. Two strategies have been pursued. The first exploits the short-range nature of dynamical correlation through localized orbital domains, as in pair-natural orbital (PNO)~\cite{Neese2009b,Schmitz2014,Pinski2015}, domain-based local PNO (DLPNO)~\cite{Riplinger2013,Riplinger2015,Riplinger2016}, and orbital-specific virtuals (OSV)~\cite{Yang2012} approaches.

The second strategy, low-rank tensor decomposition, targets the algebraic structure of the ERI tensor. Density fitting [DF, also called resolution of identity (RI)]~\cite{Dunlap1979,Feyereisen1993} and Cholesky decomposition (CD)~\cite{Beebe1977,Roeggen2008} factorize the four-index ERI tensor into two third-order tensors. Both approaches reduce the prefactor but leave the formal $O(N^5)$ contraction scaling unchanged. Pseudospectral (PS) methods~\cite{Neese2009,Friesner1986} exploit a real-space grid $\{r_g\}$ to separate the bra and ket indices as $(\mu\nu|\lambda\sigma) \approx \sum_g X_{\mu g}X_{\nu g} V_{\lambda\sigma}^g$, reducing the scaling of exchange-like contractions but not the overall tensor contraction complexity. Tensor hyper-contraction (THC)~\cite{Hohenstein2012a,Parrish2012,Lee2020} goes further by factorizing the ERI into a product of five matrices involving only two-index quantities, using real-space interpolative grids $\{r_K\}$:
\begin{equation}
(\mu\nu|\lambda\sigma) \approx \sum_{KL} X_{\mu K} X_{\nu K} Z_{KL} X_{\lambda L} X_{\sigma L},
\end{equation}
where $X_{\mu K}=\chi_\mu(r_K)\sqrt{w_K}$ and $Z_{KL}$ is the THC kernel. THC reduces the contraction scaling to $O(N^4)$. Several cubic-scaling RPA implementations exploit low-rank structure and orbital locality~\cite{Moussa2014,Duchemin2019,Wilhelm2016,Schurkus2016}, THC with $k$-point sampling enables $O(N^3)$ RPA and $GW$ for periodic systems~\cite{Yeh2023,Yeh2024}; BTD-based $GW$ achieves $O(N^3)$ scaling for molecules~\cite{Zhang2026}. For molecules, however, constructing the THC kernel itself scales as $O(N^4)$, creating a preprocessing bottleneck. The scaled opposite-spin MP2 (SOS-MP2) method~\cite{Jung2004}, which retains only the Coulomb (J) term, can be combined with our recently developed block tensor decomposition (BTD) algorithm~\cite{Zhang2025} to achieve $O(N^3)$ approximate MP2. BTD employs a dual-grid scheme based on Hilbert space-filling curves and pivoted Cholesky decomposition~\cite{Beebe1977} to construct the THC half-kernel $B_{MK}$ at formal $O(N^3)$ cost. However, the exchange (K) part of MP2 cannot be handled by THC with the same efficiency, because the orbital indices in $(ib|ja)$ are coupled across different particles, preventing a simple factorization with THC or RI.

Stochastic RI methods~\cite{Takeshita2017,Dou2019,Zhao2024} bypass the K-part bottleneck through random orbital decomposition, at the cost of statistical noise. Canonical polyadic decomposition (CPD)~\cite{Hitchcock1927,Benedikt2011} provides a fully deterministic alternative by factorizing two-electron integrals with independent factor matrices for each orbital index. In the MO basis, the CPD of $(ia|jb)$ reads
\begin{equation}
(ia|jb) \approx \sum_{r} L_{ir}L_{ar}U_{j r}U_{b r},
\end{equation}
where $L$ and $U$ each consist of two sub-matrices ($L^{\text{occ}}$, $L^{\text{vir}}$ of sizes $N_{\text{occ}}\times R$ and $N_{\text{vir}}\times R$, respectively, and analogously $U^{\text{occ}}$, $U^{\text{vir}}$), which together decouple all four orbital indices. The bra-side indices $(i,a)$ thus factor into independent occupied and virtual contributions. This fully decouples occupied and virtual indices, enabling efficient exchange contractions. Pierce and Morales~\cite{Pierce2025} proposed an $O(N^3)$ Laplace-transformed MP2 method combining THC and CPD, applying CPD exclusively to the exchange channel. Their approach differs from ours in three key respects. First, they used ISDF-based THC~\cite{Hu2017,Qin2023} whose kernel construction scales as $O(N^4)$, whereas we replace it with BTD achieving $O(N^3)$. Second, they employed a matrix-free ALS solver for the CPD, whereas we develop a block-based two-stage ALS that coarse-grains and then polishes. Third, their work is limited to MP2, whereas we extend the THC+CPD framework to RPA+SOSEX+rSE. Because of this $O(N^4)$ kernel bottleneck, a fully $O(N^3)$ molecular MP2 was not yet realized in Ref.~\cite{Pierce2025}.

Here we combine the $O(N^3)$ BTD kernel with CPD for the exchange channel, yielding a fully $O(N^3)$ BTD-MP2, and extend the framework to the complete rPT2 method (RPA + SOSEX + rSE). The principal contributions are:

\begin{enumerate}
\item[\textbf{(i)}] BTD and CPD are combined, and a block-based two-stage alternating least squares (ALS) algorithm is developed to efficiently compute the CPD factor vectors: a coarse phase solves block-diagonal subproblems in parallel, followed by a polishing phase using the full Gram matrix.
\item[\textbf{(ii)}] BTD is extended to dynamic Coulomb screening through an asymmetric half-kernel construction: the bare kernel $B$ acts on the $L$-side vertex (orbital indices $i,a$) while a coupling-constant-averaged (AC) screened kernel $\tilde{B}(i\omega)=\Pi^{\mathrm{ac}}(i\omega)\cdot B$ acts on the $U$-side vertex (indices $j,b$), enabling efficient SOSEX evaluation.
\item[\textbf{(iii)}] The rSE correction, whose exact-exchange matrix is evaluated at $O(N^3)$ cost via the chain-of-spheres exchange (COSX) algorithm~\cite{Neese2009}, is combined with BTD-RPA and the BTD-CPD SOSEX developed here. These three components are integrated within a unified asymmetric-half-kernel framework, yielding the complete BTD-rPT2 method without any single step exceeding $O(N^3)$ scaling.
\end{enumerate}

The combined BTD-rPT2 method achieves formal $O(N^3)$ computational and $O(N^2)$ storage scaling. The accuracy of the BTD approximation is validated against canonical RI-MP2; the BTD-RPA component was benchmarked in our previous work;~\cite{Zhang2025}, and the present rPT2 results are assessed on the S66x8 data set against CCSD(T)/CBS reference values.

Section~II presents the methodology. Key notation is collected in Table~I; standard quantum-chemistry indices ($i,j$ occupied, $a,b$ virtual, $\mu,\nu$ AO, $M,N$ auxiliary) are used throughout. Computational details are given at the beginning of Sec.~III, which then validates BTD-MP2, benchmarks the scaling of BTD-rPT2, and assesses accuracy via potential energy curves and the S66x8 benchmark.

\begin{table}[tb]
\caption{Notation. Standard quantum-chemistry indices are used throughout.}
\label{tab:notation}
\begin{ruledtabular}
\begin{tabular}{ll}
$i,j,k,l$ & Occupied molecular orbitals\\
$a,b,c,d$ & Virtual molecular orbitals\\
$\mu,\nu,\lambda,\sigma$ & Atomic orbital basis functions\\
$M,N$ & RI auxiliary basis functions\\
$K,L$ & BTD interpolative grid points\\
$r$ & CPD rank index\\
$\tau_t,\omega_w$ & Imaginary-time and frequency grid points\\
$X_{\mu K}$ & Real-space collocation matrix\\
$B_{MK}$ & BTD half-kernel\\
$\tilde{B}_{MK}$ & AC-screened half-kernel\\
$L_{ir},L_{ar}$ & CPD left factors (bra)\\
$U_{jr},U_{br}$ & CPD right factors (ket)\\
$\Pi^{\text{ac}}$ & Coupling-constant averaged interaction\\
$S^L_{Mr},S^U_{Mr}$ & Screened intermediates\\
$N_{\text{BTD}},N_{\text{CPD}}$ & BTD grid size and CPD rank\\
\end{tabular}
\end{ruledtabular}
\end{table}

\section{Methodology}

\subsection{Laplace-transformed RPA and rPT2}

The MP2 correlation energy in the canonical molecular orbital basis reads
\begin{equation}
E_c^{\text{MP2}} = -\frac{1}{2}\sum_{ijab}\frac{[(ia|jb) - (ib|ja)](ia|jb)}{\Delta_{ij}^{ab}},
\label{eq:mp2}
\end{equation}
where $\Delta_{ij}^{ab} = \epsilon_a + \epsilon_b - \epsilon_i - \epsilon_j$. Introducing the Laplace transform $1/x = \int_0^\infty e^{-\tau x}d\tau$~\cite{Almlof1991,Haser1992}, Eq.~(\ref{eq:mp2}) becomes an imaginary-time integral. The $\tau$ integral is discretized via minimax quadrature~\cite{Kaltak2014}, which provides an optimal set of $\{\tau_t, w_t\}$ points that minimize the maximum quadrature error. The corresponding imaginary-frequency grid $\{i\omega_w\}$ for the RPA and SOSEX frequency integration is obtained from a cosine transform of the $\tau$ grid; both the $\tau \to \omega$ transform and the AC-SOSEX $\lambda$ integration are implemented using the time-frequency component of the GreenX library~\cite{Azizi2024}:
\begin{multline}
E_c^{\text{MP2}} = -\frac{1}{2}\sum_{ijab}[(ia|jb) - (ib|ja)](ia|jb) \times\\
\int_0^\infty G_a(\tau)G_b(\tau)G_i(-\tau)G_j(-\tau)\,d\tau,
\label{eq:lt-mp2}
\end{multline}
where $G_p(\tau) = e^{-\tau\epsilon_p}$ is the imaginary-time Green's function. The Coulomb-only J-part involves only ring diagrams:
\begin{equation}
E^{\text{MP2,J}} = -\frac{1}{2}\sum_{ijab}\frac{(ia|jb)^2}{\Delta_{ij}^{ab}}.
\label{eq:jpart}
\end{equation}
The non-interacting density-density response function in imaginary time is defined as
\begin{equation}
P^0(\mathbf{r},\mathbf{r}',\tau) = \sum_{ia}\phi_i(\mathbf{r})\phi_a(\mathbf{r})P_{ia}(\tau)\phi_a(\mathbf{r}')\phi_i(\mathbf{r}').
\label{eq:p0-tau}
\end{equation}
where $P_{ia}(\tau) = G_i(-\tau)G_a(\tau)$. Applying the cosine transform $P^0(i\omega)=\int_0^\infty P^0(\tau)\cos(\omega\tau)d\tau$ yields the matrix elements of $P^0(i\omega)$ in the occupied-virtual basis:
\begin{equation}
    P_{ia}(i\omega)=\frac{2(\epsilon_i-\epsilon_a)}{(\epsilon_i-\epsilon_a)^2+\omega^2}.
\end{equation}
The sign convention $(\epsilon_i-\epsilon_a)<0$ follows from the definition $P_{ia}(\tau)=G_i(-\tau)G_a(\tau)$; most RPA literature uses the opposite sign, which is equivalent since $P_{ia}$ appears pairwise in all energy expressions.
The J-part in MO representation becomes
\begin{equation}
E^{\text{MP2,J}} = -\frac{1}{4\pi}\int_0^\infty d\omega\sum_{ia,jb}(ia|jb)^2 P_{ia}(i\omega)P_{jb}(i\omega),
\label{eq:jpart-freq}
\end{equation}
which is the second-order expansion of the RPA correlation energy:~\cite{Furche2001,Eshuis2012}
\begin{equation}
E_c^{\text{RPA}} = \frac{1}{2\pi}\int_0^\infty d\omega\,\mathrm{Tr}\bigl[\ln(1-P^0(i\omega)v) + P^0(i\omega)v\bigr],
\label{eq:rpa}
\end{equation}
where the integral runs over imaginary frequencies $i\omega$, $\ln$ denotes the matrix logarithm, and $\mathrm{Tr}$ the matrix trace. In practice, $v$ and $v^{1/2}$ are represented in the RI auxiliary basis through the Coulomb metric $V_{MN}=(M|N)$ and its inverse square root $V^{-1/2}$. RI-RPA~\cite{Ren2012b} scales as $O(N^4)$.

Alternatively, RPA can be formulated in the coupled-cluster framework as direct ring coupled-cluster doubles (drCCD)~\cite{Scuseria2008,Scuseria2013,Bleiziffer2013} where $E_c^{\text{RPA}} = \frac{1}{2}\sum_{ijab}(ia|jb)T_{ij}^{ab}$ and $T_{ij}^{ab}$ are the drCCD amplitudes obtained by solving the ring coupled-cluster doubles equations. The SOSEX correction antisymmetrizes the Coulomb integral:~\cite{Freeman1977,Gruneis2009}
\begin{equation}
E_c^{\text{RPA+SOSEX}} = \frac{1}{2}\sum_{ijab}[(ia|jb)-(ib|ja)]T_{ij}^{ab}.
\end{equation}

In the ACFDT framework, the SOSEX correction is given by the AC-SOSEX expression.~\cite{Jansen2010,Ren2013} The $\lambda$-dependent RPA screened interaction is $W_\lambda(i\omega) = \lambda v/(1-\lambda P^0(i\omega)v)$, and the coupling-constant-averaged screened interaction is $\bar{W}(i\omega) = \int_0^1 d\lambda\,W_\lambda(i\omega)$. The AC-SOSEX energy reads
\begin{widetext}
\begin{equation}
E_c^{\text{AC-SOSEX}} = -\frac{1}{2\pi}\int_0^\infty d\omega
\sum_{ia,jb}(ib|ja)\bar{W}_{ia,jb}(i\omega)P_{ia}(i\omega)P_{jb}(i\omega)
\label{eq:ac-sosex}
\end{equation}
\end{widetext}
AC-SOSEX differs negligibly from the drCCD-based SOSEX (relative error $< 0.15\%$).~\cite{Jansen2010}

For non-Hartree--Fock reference determinants, single excitations contribute at first order because Brillouin's theorem does not hold. The rSE correction~\cite{Ren2011,Ren2013} is obtained by semi-canonicalization: diagonalizing the occupied and virtual blocks of the Fock matrix $f_{pq} = \langle\psi_p|\hat{f}|\psi_q\rangle$ (evaluated with KS orbitals) yields transformed eigenvalues $\tilde{\epsilon}_i,\tilde{\epsilon}_a$ and off-diagonal elements $\tilde{f}_{ia}$. The rSE energy is then
\begin{equation}
E_c^{\text{rSE}} = 2\sum_{ia}\frac{|\tilde{f}_{ia}|^2}{\tilde{\epsilon}_a-\tilde{\epsilon}_i},
\label{eq:rse}
\end{equation}
where the factor of 2 accounts for spin summation. The complete rPT2 correlation energy is $E_c^{\text{rPT2}} = E_c^{\text{RPA}} + E_c^{\text{SOSEX}} + E_c^{\text{rSE}}$.

\subsection{BTD-CPD: BTD with canonical polyadic decomposition}

The THC format factorizes the four-index ERI tensor using real-space interpolative grids $\{r_K\}$:
\begin{equation}
(\mu\nu|\lambda\sigma) \approx \sum_{KL} X_{\mu K}X_{\nu K}Z_{KL}X_{\lambda L}X_{\sigma L},
\label{eq:thc}
\end{equation}
with $X_{\mu K} = \chi_\mu(r_K)\sqrt{w_K}$ and the THC kernel $Z_{KL}$. For molecular systems with atom-centered basis functions, constructing $Z_{KL}$ scales as $O(N^4)$. BTD~\cite{Zhang2025} overcomes this bottleneck via a dual-grid scheme. Starting from a dense set of Lebedev integration grids $\{r_g,w_g\}$, the method first constructs a set of candidate interpolative points by partitioning the dense grids into blocks along a Hilbert space-filling curve. Each block contributes a weighted centroid $r_K = \sum_{g\in K}r_g w_g/\sum_{g\in K}w_g$, where the index $K$ labels the resulting interpolative grid points (replacing the $\bar{g}$ notation used earlier). From these candidates, a pivoted Cholesky decomposition~\cite{Beebe1977,Matthews2020} of the squared overlap matrix $S_{KL} = (\sum_\mu\psi_\mu(r_K)\psi_\mu(r_L)\sqrt{w_Kw_L})^2$ with cutoff $\varepsilon^2\cdot\max(S_{KK})$ selects a compact, non-redundant subset. The BTD half-kernel $B_{MK}$, which maps auxiliary basis functions $M$ to interpolative grids $K$, is then built from 2c1e integrals and regularized via the RI Coulomb metric:
\begin{subequations}
\begin{align}
& B^\mathrm{pre}_{MK} = \sum_g (M|r_g)S_{gK},\\
& B_{MK} = \sum_{L}S^{-1}_{KL}\sum_{N}B^\mathrm{pre}_{NL}V^{-1/2}_{NM},
\label{eq:btd-kernel}
\end{align}
\end{subequations}
where $S_{gK} = \sum_{\mu} \psi_\mu(r_g)\sqrt{w_g}\,\psi_\mu(r_K)\sqrt{w_K}$ is the overlap fitting matrix between dense grid $g$ and interpolative grid $K$, The THC kernel follows as $Z_{KL} = \sum_M B_{MK}B_{ML}$, completing the $O(N^3)$ construction. Using the sparsity of $S_{gK}$, the computational cost can be further reduced.

To handle the exchange channel, where THC cannot decouple the cross-particle orbital indices in $(ib|ja)$, we employ CPD~\cite{Hitchcock1927,Benedikt2011}. CPD factorizes the MO-transformed integrals $(ia|jb)$ with independent factor matrices for each orbital index:
\begin{equation}
(ia|jb) \approx \sum_{r} L_{ir}L_{ar}U_{j r}U_{b r},
\label{eq:cpd}
\end{equation}
where $L$ and $U$ carry the bra-side $(i,a)$ and ket-side $(j,b)$ indices, respectively (Table~\ref{tab:notation}). In practice, the CPD is fitted to the BTD-transformed three-center integrals $B_{M,ia} = \sum_K B_{MK} X_{iK} X_{aK}$ rather than to the raw $(ia|jb)$; the ALS optimization therefore operates on the combined $(M,r)$ index space, where $M$ is the auxiliary-function index of the BTD kernel and $r$ is the CPD rank. This space is partitioned into blocks in a two-stage strategy. In the coarse stage, the Gram matrix and the matricized tensor times Khatri-Rao product (MTTKRP) are computed block-diagonally, each block corresponding to a subset of $r$ and yielding an independent low-dimensional Gram system that avoids the $O(N_{\text{CPD}}N_\text{BTD}^2)$ cost of a full linear solve. The formal scaling is reduced to $O(N_{\text{CPD}}N_\text{BTD}N_\text{vir})$. A final polishing stage solves the full Gram system to reach the globally optimal factorization.

For MP2, a robust CPD correction~\cite{Pierce2025} is applied to the exchange energy. The main term $E^{\text{main}}_K$ is evaluated by contracting the CPD-compressed intermediates through the BTD kernel $B_{MK}$, while the correction term $E^{\text{corr}}_K$ is evaluated by contracting the CPD factors directly---i.e., treating $B_{MK}$ as $\delta_{MK}$ in the exchange contraction, which projects the CPD representation onto the grid without the BTD interpolation. The final energy $E^{\text{final}}_K = 2E^{\text{main}}_K - E^{\text{corr}}_K$ cancels the leading-order CPD approximation error.

\subsection{BTD-rPT2: $O(N^3)$ rPT2 implementation}

We now describe how the BTD-CPD framework is applied to each component of rPT2.

The RPA (J-part) has already been demonstrated at $O(N^3)$ in our previous BTD work.~\cite{Zhang2025} We focus on the two remaining rPT2 components.

The SOSEX (K-part) motivates the asymmetric half-kernel design. Constructing the frequency-dependent THC kernel $Z_{KL}(i\omega_w)$ at each quadrature point would be prohibitively expensive. We avoid this entirely by leveraging a key property of BTD: the bare Coulomb half-kernel $B_{MK}$ is frequency-independent and needs to be constructed only once. The frequency dependence of the screening is carried entirely by the coupling-constant-averaged interaction
\begin{equation}
\Pi^{\text{ac}}(i\omega_w) = \int_0^1 d\lambda\,\lambda\,v^{1/2}\bigl[1-\lambda P^0(i\omega_w)\bigr]^{-1}v^{1/2},
\label{eq:pi-ac}
\end{equation}
which is computed at each frequency point via 7-point Gauss--Legendre quadrature~\cite{Ren2013} and applied as a single matrix multiplication $\tilde{B}_{MK}(i\omega_w) = \sum_N \Pi^{\text{ac}}_{MN}(i\omega_w) B_{NK}$. This costs only $O(N_{\text{aux}}^2 N_\text{BTD})$ per frequency, negligible compared to the CPD optimization.
Since $\Pi^{\text{ac}}(i\omega_w)$ is symmetric, a symmetric half-kernel $\bar{B} = (\Pi^{\text{ac}})^{1/2} B$ is formally equivalent, but constructing $(\Pi^{\text{ac}})^{1/2}$ via SVD is ill-conditioned due to the rank deficiency of $\Pi^{\text{ac}}$. The asymmetric form $\tilde{B} = \Pi^{\text{ac}} B$ with bare $B$ avoids this decomposition while delivering identical physics.

With the half-kernels defined, the evaluation proceeds as follows. The imaginary time $\{\tau_t\}$ and frequency $\{i\omega_w\}$ grids follow the minimax quadrature of Kaltak et al.~\cite{Kaltak2014} standard in $GW$ and RPA. The CPD factorization of Eq.~(\ref{eq:cpd}) is performed once before the frequency loop. The collocation matrix is transformed to the MO basis as $X_{pK} = \sum_\mu C_{\mu p}X_{\mu K}$, where $C_{\mu p}$ are MO coefficients. The CPD-compressed Green's function products on the BTD grids are then
\begin{subequations}
\begin{align}
G^{\text{occ},L}_{Kr}(\tau_t) &= \sum_i X_{iK}G_i(\tau_t)L_{ir},\\
G^{\text{vir},L}_{Kr}(\tau_t) &= \sum_a X_{aK}G_a(\tau_t)L_{ar},
\label{eq:sosex-g}
\end{align}
\end{subequations}
with analogous expressions for $G^{\text{occ},U},G^{\text{vir},U}$ using $U_{jr},\ U_{br}$. The particle-hole products are transformed to the frequency domain via a cosine transform weighted by the time quadrature weights $w_t$:
$Q^L_{Kr}(i\omega_w) = \sum_t w_t \cos(\omega_w \tau_t) G^{\text{occ},L}_{Kr}(\tau_t) G^{\text{vir},U}_{Kr}(\tau_t)$,
and $Q^U_{Kr}(i\omega_w)$ analogously with $G^{\text{occ},U}$ and $G^{\text{vir},L}$. The cross-pairing of $L$ and $U$ factor groups mirrors the exchange contraction in $(ib|ja)$, where the two occupied (and two virtual) indices belong to different CPD factor sets. The asymmetric screening is then applied:
\begin{subequations}
\begin{align}
&S^L_{Mr}(i\omega_w) = \sum_K B_{MK}Q^L_{Kr}(i\omega_w),\\
&S^U_{Mr}(i\omega_w) = \sum_K \tilde{B}_{MK}(i\omega_w)Q^U_{Kr}(i\omega_w),
\label{eq:sosex-screen}
\end{align}
\end{subequations}
where $\tilde{B}_{MK}(i\omega_w) = \sum_N \Pi^{\text{ac}}_{MN}(i\omega_w)B_{NK}$. The SOSEX energy is assembled as
\begin{equation}
E_c^{\text{SOSEX}} = \frac{1}{\pi}\sum_w\omega_w\sum_{M,r}S^L_{Mr}(i\omega_w)S^U_{Mr}(i\omega_w).
\label{eq:sosex-energy}
\end{equation}
Here $\omega_w$ denotes the combined quadrature weight and normalization factor; the apparent factor of 2 difference from the $1/(2\pi)$ prefactor in Eq.~(\ref{eq:ac-sosex}) arises from two sources: the negative sign of the exchange integral is absorbed into the $S^L S^U$ pairing, and the $2/\pi$ factor from Parseval's relation for the cosine transform is distributed across the $\omega_w$ weights and the $1/\pi$ prefactor; the two expressions are numerically equivalent.

Unlike the MP2 K-part described in Sec.~II.B, no robust CPD correction is applied here. A frequency-dependent CPD of $\bar{W}(i\omega)$ would be prohibitively expensive---a single CPD optimization already costs $O(N_\text{CPD}N_{\text{BTD}}^2)$ per ALS sweep, and repeating it for $n_\omega \sim 16$ frequency points would erase all scaling gains. The resulting CPD error in the SOSEX term is expected to be smaller than in bare MP2 exchange because the screened interaction $\bar{W}$ already incorporates correlation effects that suppress high-energy orbital pairs. The smooth, noise-free potential energy curves (Sec.~III.D) further confirm the numerical stability of the BTD dual-grid scheme.

Finally, the rSE correction of Eq.~(\ref{eq:rse}) requires the exact-exchange matrix, which is obtained at $O(N^3)$ cost via the COSX algorithm.~\cite{Neese2009} Semi-canonicalization---diagonalizing the $f_{ij}$ and $f_{ab}$ blocks of the Fock matrix---yields $\tilde{f}_{ia}$ and the transformed orbital energies $\tilde{\epsilon}_i,\tilde{\epsilon}_a$. The total rPT2 energy is $E_c^{\text{rPT2}} = E_c^{\text{RPA}} + E_c^{\text{SOSEX}} + E_c^{\text{rSE}}$.

Table~\ref{tab:scaling} summarizes the formal scaling of each computational step. The overall formal scaling is $O(N^3)$, dominated by the BTD kernel construction, the RPA polarization build, and the COSX exchange. The SOSEX-specific steps scale as $O(N^3)$ or lower owing to the CPD compression.

\begin{table}[tb]
\caption{Formal scaling of the computational steps in BTD-rPT2. $N$ denotes system size; $n_\tau$ and $n_\omega$ (imaginary-time and frequency grid points) are system-independent. All other quantities scale linearly with $N$.}
\label{tab:scaling}
\begin{ruledtabular}
\begin{tabular}{ll}
\textbf{Step} & \textbf{Scaling}\\
\hline
BTD kernel construction (once) & $O(N^3)$\\
CPD factorization (once) & $O(N N_{\text{CPD}} N_\text{BTD})$\\
RPA polarization $\Pi^0_{MN}(i\omega_w)$ & $O(N^3 n_\omega)$\\
AC screening $\Pi^{\text{ac}}(i\omega_w)$ & $O(N^3 n_\omega)$\\
CPD Green's function products & $O(N^2 N_{\text{CPD}} n_\tau)$\\
SOSEX energy assembly & $O(N^2 n_\omega)$\\
RPA energy evaluation & $O(N_{\text{aux}}^3 n_\omega)$\\
rSE correction & $O(N^3)$
\end{tabular}
\end{ruledtabular}
\end{table}

\section{Results and Discussion}

\subsection{Computational details}

All calculations were performed with a development version of the {\sc XEDA} package. The BTD half-kernel was constructed using $N_{\text{BTD}} \approx 5 N_{\text{aux}}$ interpolative grid points with a pivoted Cholesky cutoff of $5\times10^{-5}$. The CPD rank was set to $N_{\text{CPD}} = 3.5\,N_{\text{occ}}$, ensuring $N_{\text{CPD}} \propto N$ as required for formal $O(N^3)$ scaling; the block-based two-stage ALS used 10 coarse iterations and 2 polishing iterations. Imaginary-time and frequency grids employed 12 minimax quadrature points.~\cite{Kaltak2014} The rSE correction was computed via COSX with the default grid. Calculations were performed on Intel Xeon Gold 6130 2.10GHz with 32 threads.

Scaling benchmarks (Sec.~III.B--III.C) used the def2-TZVPP basis with the def2-TZVPP-RI auxiliary basis~\cite{Weigend2002} (def2-TZVP/def2-TZVP-RI for BTD-MP2 validation), an HF reference, and a single CPU core. The S66x8 benchmarks (Sec.~III.E) used aug-cc-pVDZ with counterpoise correction, testing both HF and PBE0 references. Benzene potential energy curves (Sec.~III.D) were computed at aug-cc-pVDZ and aug-cc-pVTZ levels.

\subsection{Validation of BTD-MP2}

\begin{figure}[tb]
\centering
\includegraphics[width=\columnwidth]{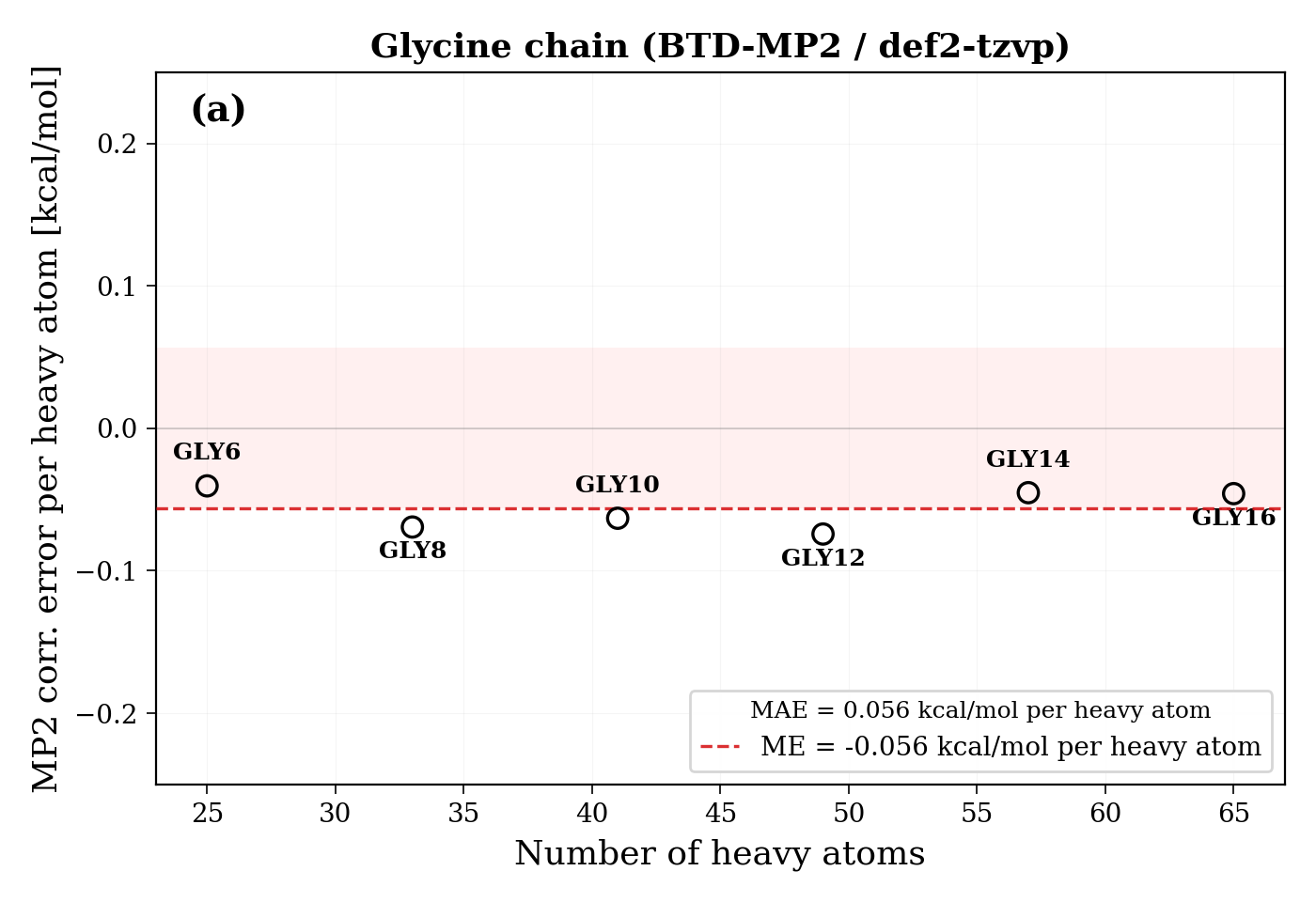}\\
\includegraphics[width=\columnwidth]{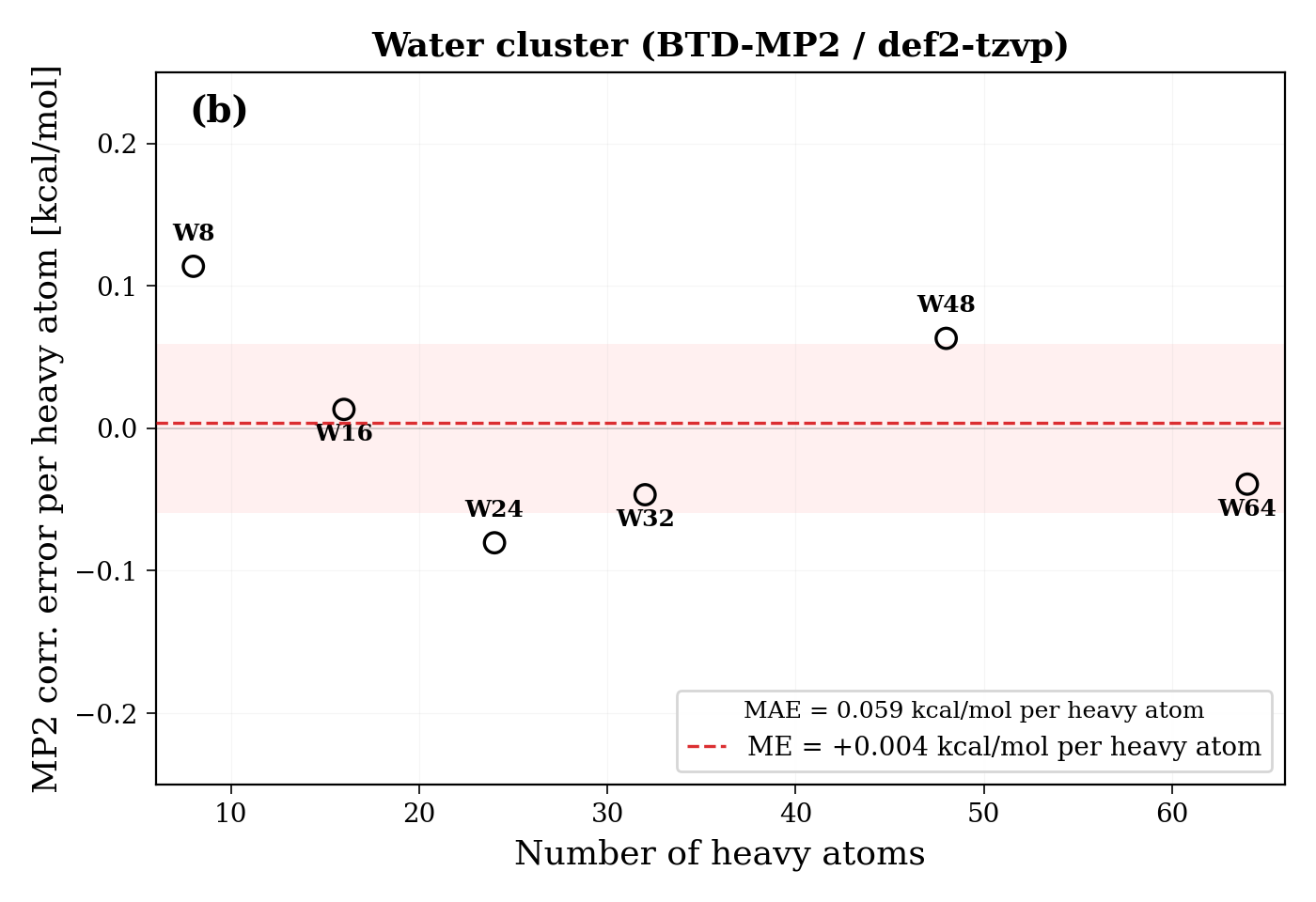}
\caption{Per-heavy-atom error of BTD-MP2 relative to canonical RI-MP2 (def2-TZVP, kcal/mol). (a) Glycine chains, (b) water clusters. Red dashed line: mean error (ME).}
\label{fig:mp2_error}
\end{figure}

We first verify that the BTD-CPD framework reproduces the canonical RI-MP2 result. In BTD-MP2, the Coulomb (J) channel is handled by BTD and the exchange (K) channel by CPD with the robust correction of Sec.~II.B. Figure~\ref{fig:mp2_error} shows the per-heavy-atom error (BTD-MP2 $-$ RI-MP2) at the def2-TZVP level. Across all 12 systems, the MAE is $0.058$~kcal/mol per heavy atom, with mean errors of $-0.056$ (glycine) and $+0.004$~kcal/mol per heavy atom (water). The largest absolute per-atom error is $0.11$~kcal/mol, corresponding to a total error of $0.91$~kcal/mol for the 8-water cluster. No systematic drift with system size is observed, confirming that the combined BTD-CPD scheme reproduces the RI-MP2 correlation energy to well below 1~kcal/mol.

The wall time of BTD-MP2 is compared against the RI-MP2 module in Psi4 in the Supporting Information (Table~S1). \cite{psi4} For the largest glycine chain (GLY$_{16}$, 2640 basis functions), BTD-MP2 completes in 1432~s versus 2081~s for RI-MP2---a $1.5\times$ speedup. For the 64-water cluster (3072 basis functions), the speedup reaches $2.4\times$. At smaller system sizes BTD-MP2 is slower than RI-MP2 due to the constant prefactor of the BTD kernel construction; the crossover occurs near 2000 basis functions. The effective scaling exponent of BTD-MP2 is $O(N^{2.60})$ for glycine and $O(N^{2.51})$ for water, well below the formal $O(N^3)$ and consistent with sub-linear growth of the BTD grid-point count.
\begin{figure*}[htbp]
\centering
\includegraphics[width=0.48\textwidth]{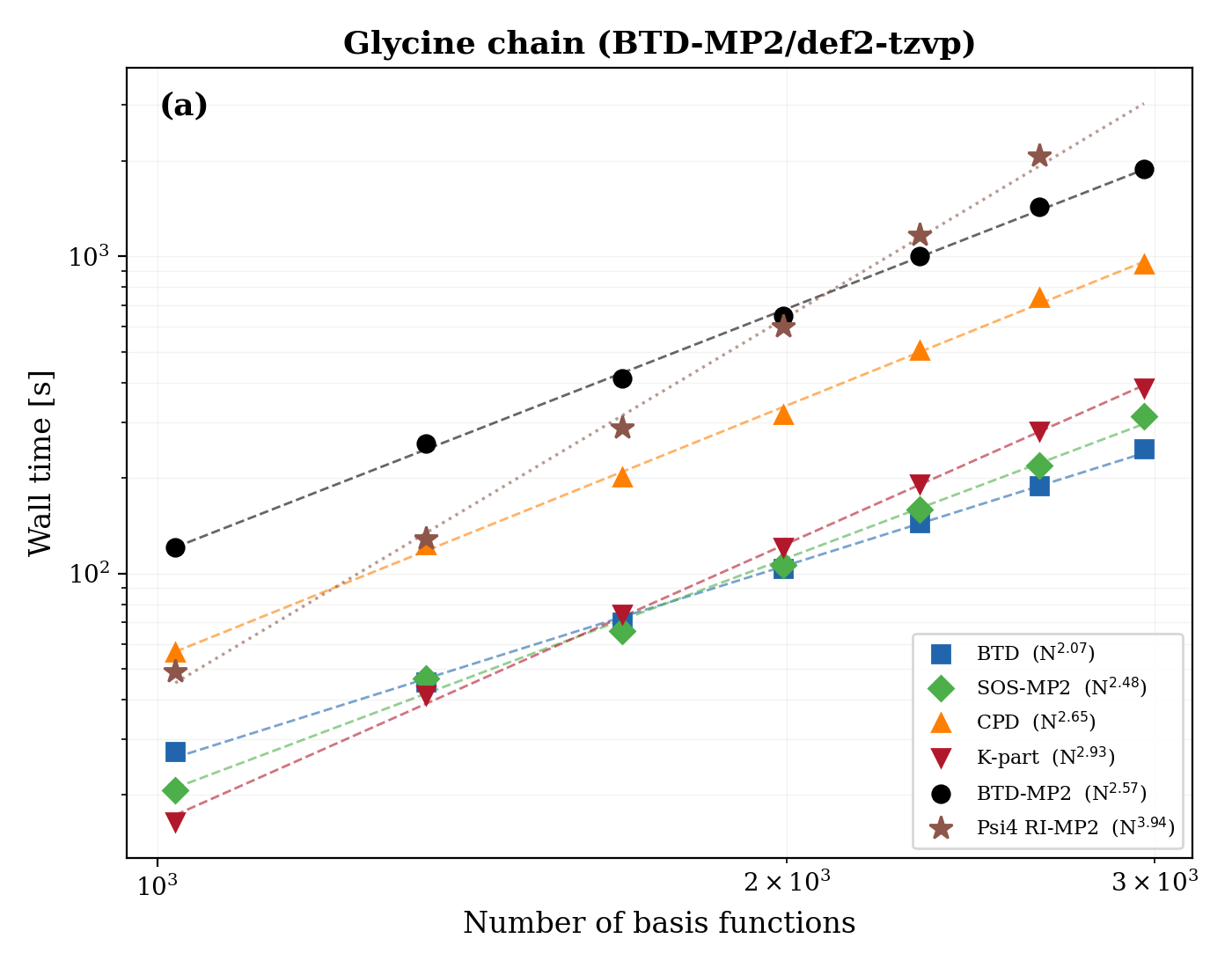}\hfill
\includegraphics[width=0.48\textwidth]{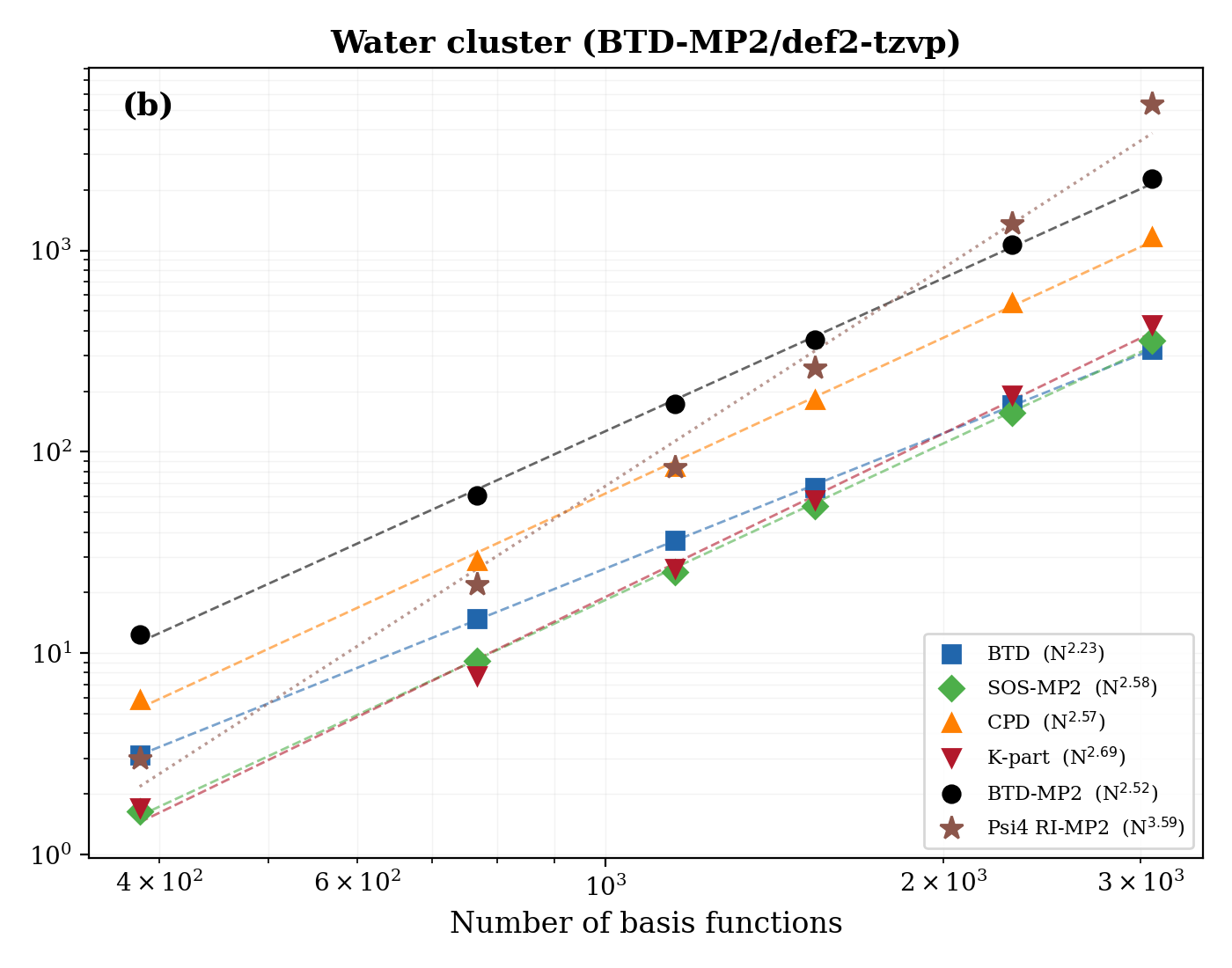}
\caption{Per-step wall-time scaling of BTD-MP2 (def2-TZVP). (a) Glycine chains, (b) water clusters. Solid lines: linear regression; dashed: $O(N^3)$ reference.}
\label{fig:S2}
\end{figure*}
Having validated BTD-MP2, we now extend the framework to the full rPT2 method. The BTD approximation (J-channel) is covered by the validation above and by previous BTD-RPA benchmarks~\cite{Zhang2025}. The CPD approximation (K-channel) is controlled by the robust correction for MP2 (Sec.~II.B); for SOSEX the CPD error is expected to be even smaller because $\bar{W}$ suppresses high-energy orbital pairs (Sec.~II.C). The rSE correction is independent of both approximations. BTD-rPT2 is therefore expected to match canonical rPT2 accuracy to within $\sim0.06$~kcal/mol per heavy atom.

\subsection{Computational scaling of BTD-rPT2}

Scaling tests used an HF reference (rSE = 0) with the def2-TZVPP basis and matching RI auxiliary basis, run on a single CPU core. Two test sets were chosen: one-dimensional glycine chains (gly$_n$, $n=6$--$16$, 45--115 atoms) and three-dimensional water clusters ($m$W, $m=8$--$48$, 24--144 atoms).

\begin{figure*}
\centering
\includegraphics[width=0.48\textwidth]{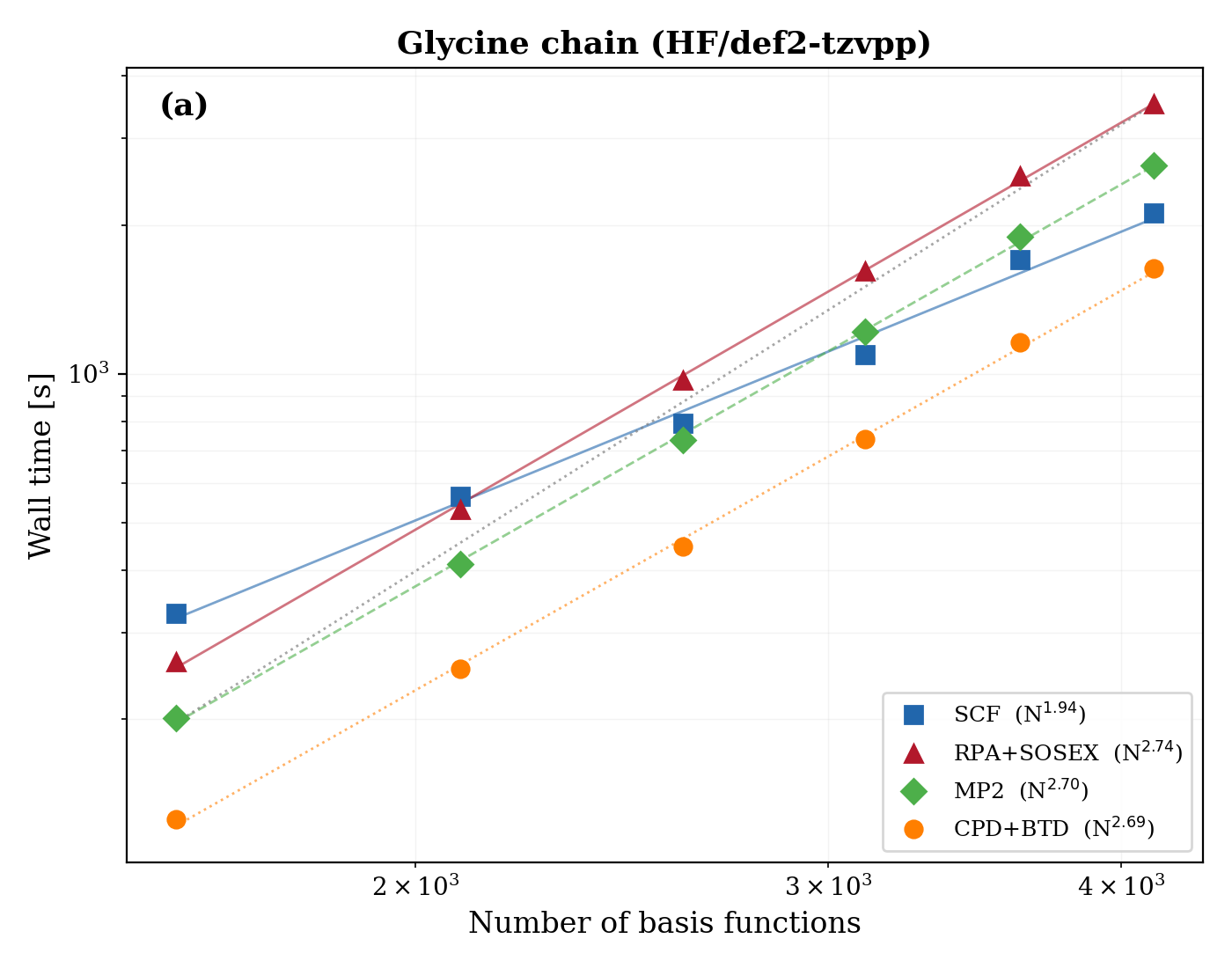}\hfill
\includegraphics[width=0.48\textwidth]{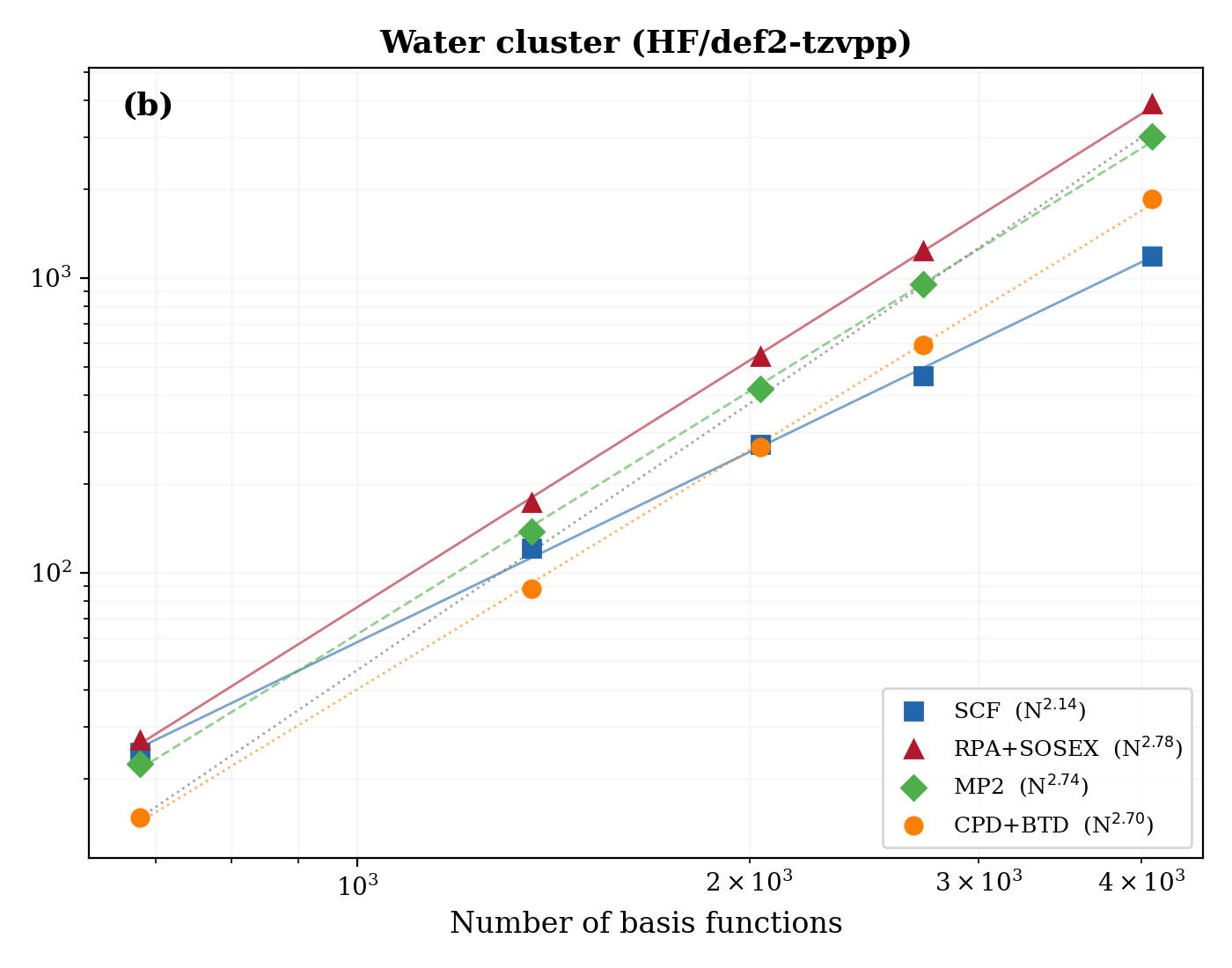}
\caption{Log--log wall-time scaling of BTD-rPT2 (def2-TZVPP, HF reference) vs.\ $N_{\text{BF}}$. (a) Glycine chains, (b) water clusters. Solid: regression fit; dashed: $O(N^3)$ reference.}
\label{fig:scaling}
\end{figure*}

Figure~\ref{fig:scaling} shows wall-time scaling versus basis-set size $N_{\text{BF}}$ for RPA+SOSEX, the MP2(J+K) component, and the CPD-BTD construction step; SCF timings are included for reference. Fitted exponents from log--log linear regression are listed in Table~\ref{tab:scaling-exponents}.

\begin{table}[tb]
\caption{Effective scaling exponents from log--log linear regression of wall time against $N_{\text{BF}}$.}
\label{tab:scaling-exponents}
\begin{ruledtabular}
\begin{tabular}{lcc}
\textbf{Component} & \textbf{Glycine (1D)} & \textbf{Water (3D)}\\
\hline
RPA+SOSEX   & 2.74 & 2.78\\
MP2    & 2.70 & 2.74\\
SCF &1.94 & 2.14\\
CPD-BTD construction & 2.77 & 2.76\\
\end{tabular}
\end{ruledtabular}
\end{table}

All correlation steps scale below the formal $O(N^3)$ ceiling (Table~\ref{tab:scaling-exponents}): RPA+SOSEX, MP2, and CPD-BTD construction exhibit exponents of 2.74, 2.70, and 2.77 for glycine, and 2.78, 2.74, and 2.76 for water. These sub-cubic values reflect the sub-linear growth of the BTD interpolative-grid count and the CPD rank with system size. Thanks to the CPD factorization, all frequency-dependent intermediates are stored as rank-$N_{\text{CPD}}$ factor matrices, yielding $O(N^2)$ storage---a decisive advantage over conventional RI methods that require $O(N^3)$ storage for the three-index polarization propagator.

The overhead of RPA+SOSEX relative to MP2 is modest: the frequency-loop screening adds $\sim30\%$ to the MP2 time. For the largest glycine chain (gly$_{16}$, 115 atoms, 4131 basis functions), the total wall time is 5633~s (94~min, SCF 38\% $+$ RPA+SOSEX 62\%). For the largest water cluster (48W, 144 atoms, 4080 basis functions), the total is 5091~s (85~min, SCF 23\% $+$ RPA+SOSEX 77\%). The one-time CPD-BTD construction dominates the correlation cost, accounting for 46--48\% of the RPA+SOSEX time in both systems. A detailed per-component breakdown is provided in the Supporting Information (Table~S4). At smaller system sizes (GLY$_6$, GLY$_8$, W8--W16), BTD-MP2 is slower than RI-MP2 due to the constant prefactor of the BTD kernel construction; the crossover occurs near 2000 basis functions. Beyond this point, the superior scaling of BTD-MP2 becomes evident (see Table~S1).

The frequency-dependent part of SOSEX---construction of $\Pi^{\text{ac}}(i\omega_w)$ and the asymmetric screening---contributes under 0.5\% of the total correlation time, so the minimax quadrature and asymmetric half-kernel design remove the frequency-domain bottleneck. The rSE correction, which vanishes for the HF reference used here, requires a single COSX exchange-matrix evaluation~\cite{Neese2009} whose $O(N^3)$ cost is negligible compared to the RPA+SOSEX frequency loop; full rPT2 scaling remains unchanged.

Having established the computational efficiency, we now turn to the accuracy of the method, beginning with a detailed analysis of potential energy curves and the influence of basis-set and reference-orbital choices.

\subsection{Potential energy curves and long-range behavior}

Potential energy curves were computed for the parallel-displaced benzene dimer across two basis sets and two reference orbitals, with intermolecular separations ranging from $0.9\times$ to $2.0\times$ the equilibrium distance.

\begin{figure*}
\centering
\includegraphics[width=0.32\textwidth]{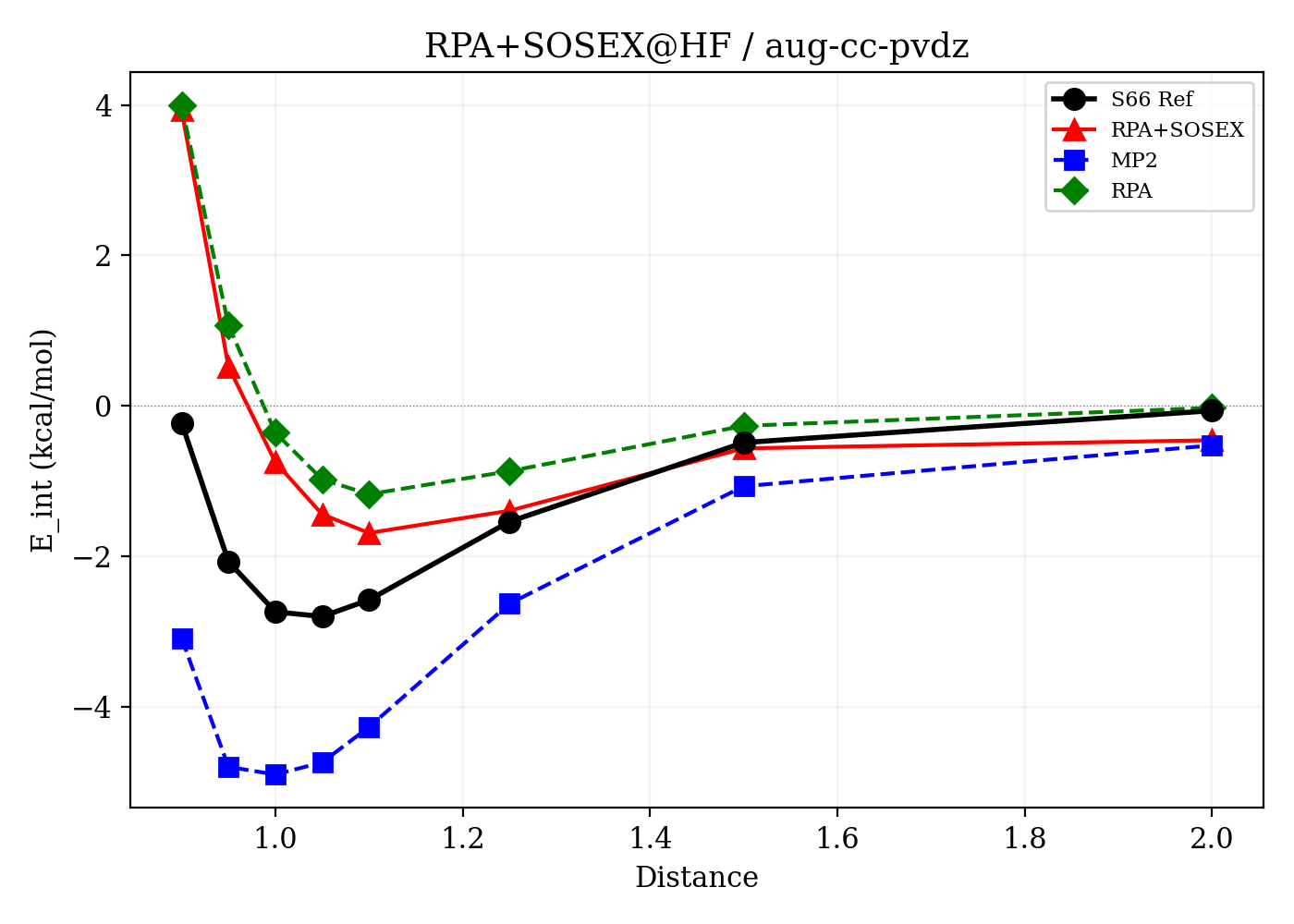}\hfill
\includegraphics[width=0.32\textwidth]{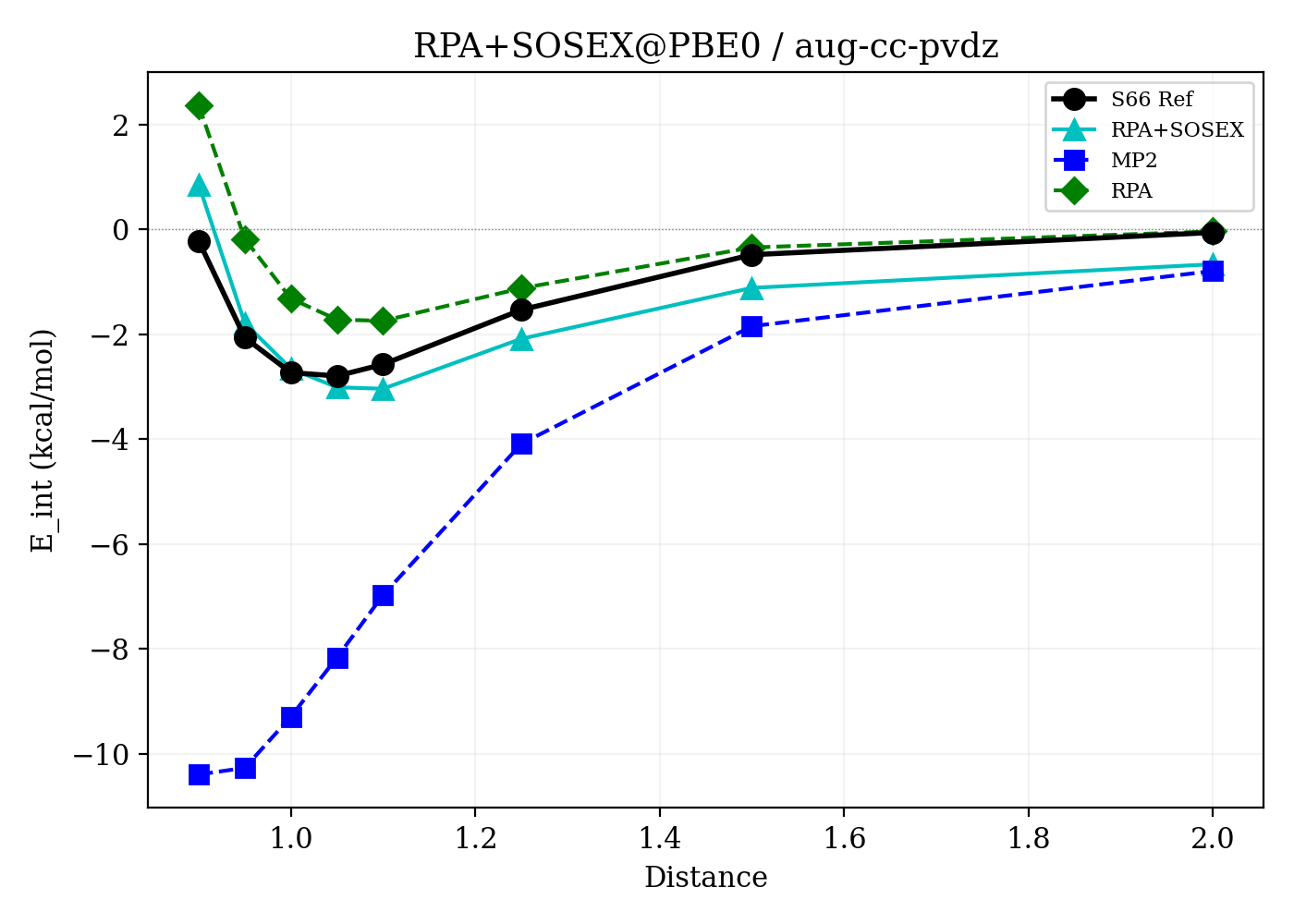}\hfill
\includegraphics[width=0.32\textwidth]{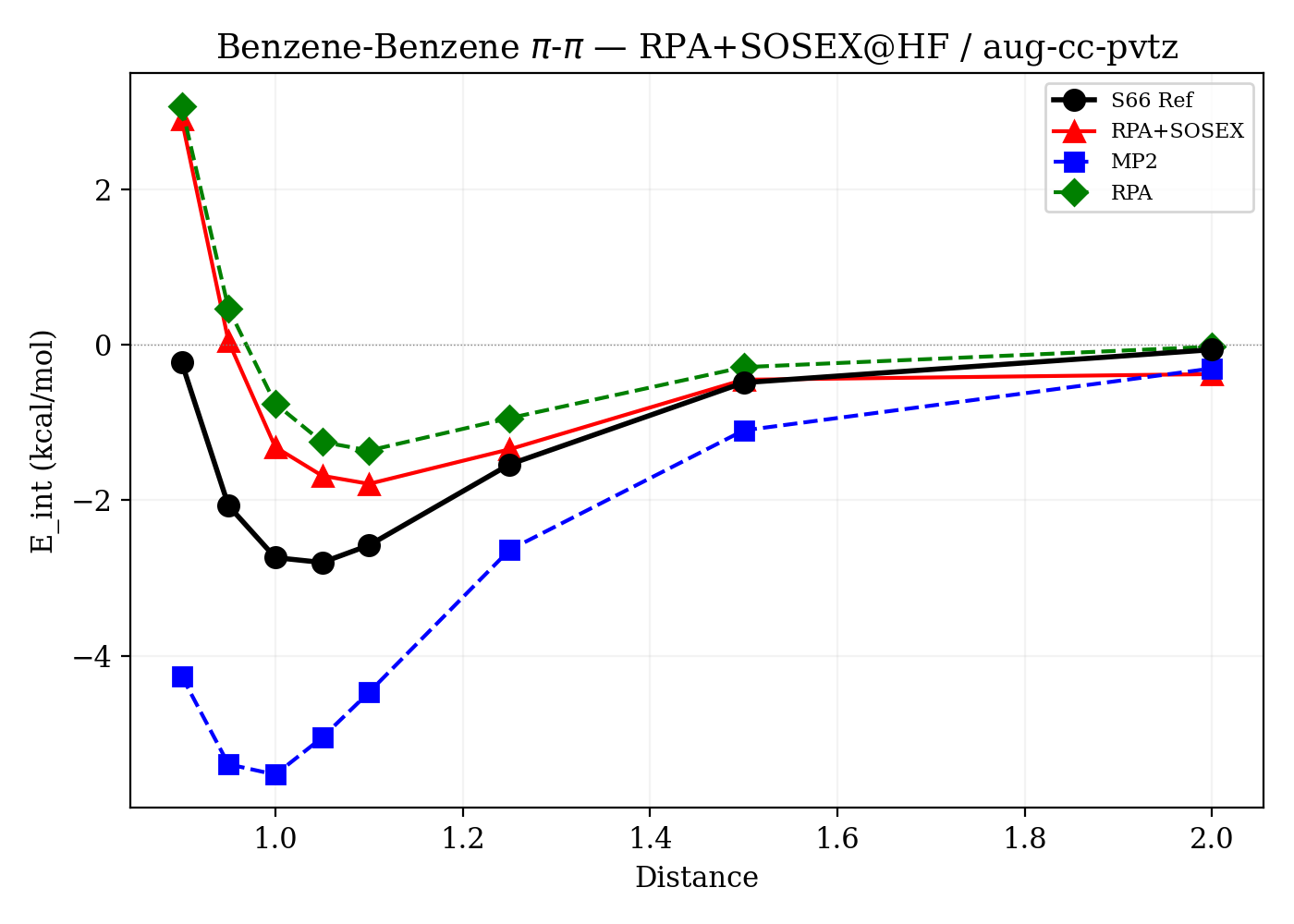}
\caption{Potential energy curves of the parallel-displaced benzene dimer. (a) HF/aug-cc-pVDZ, (b) PBE0/aug-cc-pVDZ, (c) HF/aug-cc-pVTZ. CCSD(T)/CBS reference from Ref.~\cite{Rezac2011} All CP-corrected.}
\label{fig:benzene}
\end{figure*}

Figure~\ref{fig:benzene} shows a clear pattern across basis sets and reference orbitals. At the DZ level with the HF reference [Fig.~\ref{fig:benzene}(a)], RPA tracks the CCSD(T)/CBS reference accurately across all distances, with only slight underbinding at the minimum. MP2@HF and RPA+SOSEX@HF both reproduce the well depth reasonably well but progressively overestimate the attraction at long range: at $d=2.0$, where the reference interaction is merely $-0.06$~kcal/mol, MP2@HF gives $-0.52$ and RPA+SOSEX@HF gives $-0.46$~kcal/mol. The same pattern appears with the PBE0 reference [Fig.~\ref{fig:benzene}(b)], where rPT2@PBE0 overestimates the long-range attraction even more severely ($-0.80$~kcal/mol at $d=2.0$), while RPA@PBE0 remains close to the reference. Enlarging the basis to TZ [Fig.~\ref{fig:benzene}(c)] substantially suppresses the artifact: MP2@HF/TZ drops to $-0.30$~kcal/mol at $d=2.0$, and the entire curve shifts toward the CCSD(T)/CBS result. Across all panels, RPA consistently yields the correct long-range asymptote regardless of basis set or reference orbital.

This artifact is a known consequence of the counterpoise (CP) correction in incomplete basis sets. The Boys--Bernardi scheme~\cite{Boys1970} places the partner monomer's full basis on each monomer; at large separations these ghost functions create spurious exchange stabilization. Exchange-containing correlation methods (MP2, SOSEX, rPT2) inherit this artificial attraction. RPA, by contrast, is unaffected: its correlation energy expression contains only Coulomb-type integrals $(ia|jb)$ and no exchange-type $(ib|ja)$ integrals. The DZ $\to$ TZ improvement confirms that the artifact vanishes as the basis approaches completeness.

For routine BTD-rPT2 applications with moderate basis sets, we recommend at least TZ quality or basis-set extrapolation to suppress the long-range CP artifact. Alternatively, SAPT-based analyses~\cite{Jeziorski1994} can serve as an intrinsically BSSE-free diagnostic, though their combination with BTD-rPT2 is deferred to future work. Numerical PEC data are tabulated in the Supporting Information (Tables~S5--S7).

\subsection{Accuracy on the S66x8 benchmark}

We benchmarked the accuracy on the S66x8 set~\cite{Rezac2011,Rezac2011b}, a subset of the GMTKN55 database~\cite{Goerigk2017} 66 non-covalent dimers in three categories---hydrogen-bonded (HB, 23), dispersion (DISP, 23), and mixed (MIXED, 20)---each at eight separations ($0.9\times$--$2.0\times$ equilibrium). Both HF and PBE0 reference orbitals were tested; with HF, rSE vanishes, isolating RPA+SOSEX, while with PBE0 the full rPT2 (RPA+SOSEX+rSE) is evaluated. All calculations used aug-cc-pVDZ with counterpoise correction.

\begin{figure*}
\centering
\includegraphics[width=0.85\textwidth]{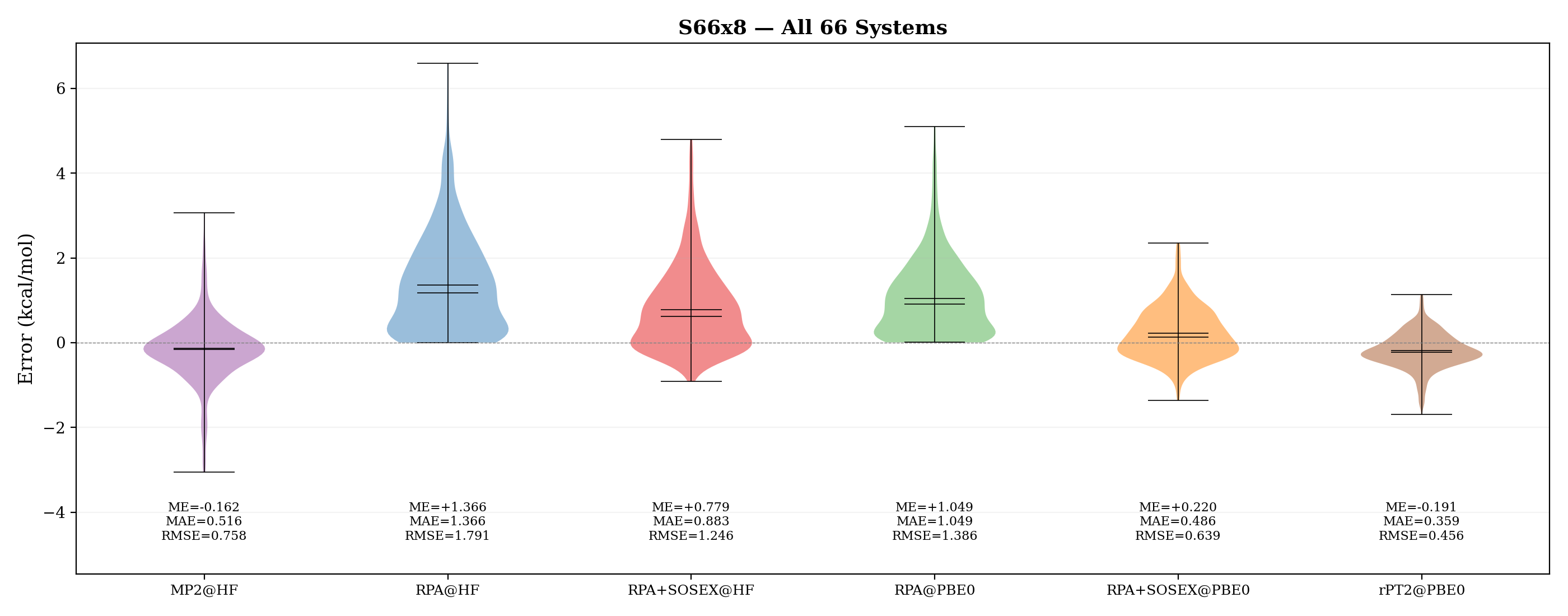}
\caption{Error distributions across all 528 data points of the S66x8 benchmark (aug-cc-pVDZ, CP-corrected). Violin plots annotated with ME, MAE, and RMSE (kcal/mol).}
\label{fig:s66x8_error}
\end{figure*}

Figure~\ref{fig:s66x8_error} and Table~\ref{tab:s66x8-summary} summarize the error statistics across all 528 S66x8 data points. At the HF reference, MP2 gives an overall MAE of 0.52~kcal/mol with a near-zero ME of $-$0.16~kcal/mol. RPA@HF systematically overbinds (ME $+$1.37, MAE 1.37~kcal/mol). Adding SOSEX reduces the MAE to 0.88~kcal/mol (ME $+$0.78), a 36\% improvement through cancellation of the same-spin self-correlation. At the PBE0 reference, the same trend holds: RPA@PBE0 overbinds (ME $+$1.05, MAE 1.05~kcal/mol), SOSEX reduces the MAE to 0.49~kcal/mol (ME $+$0.22), and the full rPT2@PBE0 achieves the best accuracy with an MAE of 0.36~kcal/mol, ME of $-$0.19~kcal/mol, and RMSE of 0.46~kcal/mol.

\begin{table}[tb]
\caption{Error statistics across all 528 S66x8 data points (aug-cc-pVDZ, CP-corrected).}
\label{tab:s66x8-summary}
\begin{ruledtabular}
\begin{tabular}{lrrr}
Method & ME & MAE & RMSE \\
\hline
MP2@HF          & $-$0.16 & 0.52 & 0.76 \\
RPA@HF          & $+$1.37 & 1.37 & 1.79 \\
RPA$+$SOSEX@HF  & $+$0.78 & 0.88 & 1.25 \\
RPA@PBE0        & $+$1.05 & 1.05 & 1.39 \\
RPA$+$SOSEX@PBE0 & $+$0.22 & 0.49 & 0.64 \\
rPT2@PBE0       & $-$0.19 & 0.36 & 0.46 \\
\end{tabular}
\end{ruledtabular}
\end{table}

The per-category breakdown (Supporting Information, Table~S9) reveals further detail. At the HF reference, MP2 gives MAEs of 0.48 (HB), 0.70 (DISP), and 0.34~kcal/mol (MIXED). RPA systematically overbinds across all categories (MAEs 1.32, 1.66, and 1.09~kcal/mol). Adding SOSEX reduces the MAE by 51\% for HB (to 0.65), 24\% for DISP (to 1.26), and 34\% for MIXED systems (to 0.72~kcal/mol). At the PBE0 reference, RPA@PBE0 gives MAEs of 1.28 (HB), 1.03 (DISP), and 0.81~kcal/mol (MIXED). RPA+SOSEX@PBE0 reduces these to 0.45, 0.61, and 0.39~kcal/mol. The full rPT2@PBE0 achieves 0.36 (HB), 0.41 (DISP), and 0.31~kcal/mol (MIXED). The rSE correction provides the largest relative gain for HB systems (20\% reduction over SOSEX@PBE0), where singles compensate for the non-Brillouin PBE0 reference. RPA overbinds most severely for dispersion-bound dimers, a known consequence of its same-spin self-correlation error.

These S66x8 results were obtained at aug-cc-pVDZ with counterpoise correction; as discussed in Sec.~III.D, exchange-containing correlation methods carry a CP-related long-range artifact at the DZ level that recedes at TZ or with basis-set extrapolation.

No standard, publicly available canonical rPT2 code exists for a direct head-to-head comparison. The BTD approximation is validated instead through three independent checks: BTD-MP2 matches RI-MP2 to $0.058$~kcal/mol per heavy atom (Sec.~III.B); BTD-RPA was previously benchmarked against RI-RPA~\cite{Zhang2025}; and the potential energy curves (Sec.~III.D) are smooth and noise-free across all separations. BTD-rPT2 thus delivers canonical rPT2 accuracy at $O(N^3)$ computational and $O(N^2)$ storage cost. Per-dimer S66x8 energies at all eight separations are in the Supporting Information (Tables~S8a, S8b, and CSV files).

\section{Conclusions}

BTD-rPT2 achieves formal $O(N^3)$-scaling renormalized second-order perturbation theory by combining block tensor decomposition with canonical polyadic decomposition. The BTD dual-grid scheme builds the THC half-kernel at $O(N^3)$ cost; CPD handles exchange through a block-based two-stage ALS; and an asymmetric half-kernel---bare Coulomb on one vertex, AC screening on the other---captures SOSEX without a frequency-dependent CPD. The rSE correction uses COSX at $O(N^3)$.

BTD-MP2 reproduces canonical RI-MP2 to $0.058$~kcal/mol per heavy atom, with wall-time scaling of $O(N^{2.5})$--$O(N^{2.8})$ on glycine chains and water clusters. On S66x8, BTD-rPT2@PBE0 gives MAE $=$ 0.36~kcal/mol (ME $-$0.19, RMSE 0.46), outperforming RPA (MAE 1.05) and RPA+SOSEX (MAE 0.49~kcal/mol). The CPD-compressed intermediates enable $O(N^2)$ storage, an advantage over conventional $O(N^3)$-storage RI-RPA.

The BTD-CPD framework is well suited to GPU acceleration, since its dominant steps---BTD kernel construction, CPD-ALS, and the frequency-loop contractions---are all dense linear algebra operations. Extension to excited-state methods such as ADC(2) is straightforward within the same Laplace-transformed framework. Finally, the long-range BSSE artifact discussed in Sec.~III.D motivates the development of a BTD-based SAPT implementation, which would provide an intrinsically BSSE-free interaction energy decomposition at $O(N^3)$ cost.

\bibliography{refs}

\end{document}